\documentclass[letter]{ieice}

\usepackage[dvips]{graphicx}
\newcommand{\mli}{\hspace{2pt}\mbox{\small $\rightarrow$}\hspace{2pt}}                      
\newcommand{\mlc}{\hspace{2pt}\mbox{\small $\wedge$}\hspace{2pt}}   
\newcommand{\add}{\hspace{2pt}\mbox{\small $\sqcup$}\hspace{2pt}}                     
\newcommand{\adc}{\hspace{2pt}\mbox{\small $\sqcap$}\hspace{2pt}} 

\newtheorem{theoremm}{Theorem}[section]
\newtheorem{factt}[theoremm]{Fact}
\newtheorem{corollaryy}[theoremm]{Corollary}
\newtheorem{definitionn}[theoremm]{Definition}
\newtheorem{thesiss}[theoremm]{Thesis}
\newtheorem{lemmaa}[theoremm]{Lemma}
\newtheorem{conventionn}[theoremm]{Convention}
\newtheorem{examplee}[theoremm]{Example}
\newtheorem{exercisee}[theoremm]{Exercise}
\newtheorem{remarkk}[theoremm]{Remark}

\newenvironment{lemma}{\begin{lemmaa}}{\end{lemmaa}}

\newenvironment{numberedlist}
{\begin{list}{\makebox[20pt]{\hss(\arabic{itemno})\enspace}}
             {\usecounter{itemno}\labelwidth 20pt}}{\end{list}}

\newcounter{itemno}

\newcounter{itemno1}

\newcounter{itemno2}

\newcounter{exno}

\newcounter{defno}

\newenvironment{defn}{\refstepcounter{defno}\medskip \noindent {\bf
Definition \thedefno.\ }}{\medskip}

\newcommand{\sep}{\;\vert\;}

\newcommand{\oprove}{\vdash\kern-.6em\lower.7ex\hbox{$\scriptstyle O$}\,}

\newcommand{\pderivation}{{\cal P}\kern -.1em\hbox{\rm -derivation}}
\newcommand{\pderivationl}{{\cal P}\kern -.1em\hbox{\em -derivation}}
\newcommand{\pderivable}{{\cal P}\kern -.1em\hbox{\rm -derivable}}
\newcommand{\pderivablel}{{\cal P}\kern -.1em\hbox{\em -derivable}}
\newcommand{\pderivations}{{\cal P}\kern -.1em\hbox{\rm -derivations}}
\newcommand{\pderivability}{{\cal P}\kern -.1em\hbox{\rm -derivability}}

\newcommand{\ie}{{\em i.e.}}

\newsavebox{\lpartfig}
\newsavebox{\rpartfig}

\newenvironment{exmple}{
 \begingroup \begin{tabbing} \hspace{2em}\= \hspace{3em}\= \hspace{3em}\=
\hspace{3em}\= \hspace{3em}\= \hspace{3em}\= \kill}{
 \end{tabbing}\endgroup}

\newcommand{\lb}{\langle}
\newcommand{\rb}{\rangle}

\newcommand{\intp}{intp_o}

\field{}

\title{Removing Qualified Names in  Modular Languages}
\authorlist{
\authorentry{Keehang Kwon}{m}{labelA}
\authorentry{Daeseong Kang}{n}{labelB}
}
\affiliate[labelA]{The author is a professor of  
Computer Eng., DongA University. email:khkwon@dau.ac.kr}
\affiliate[labelB]{The author is a professor of  
Electronics Eng., DongA University.}

\received{2003}{1}{1}
\revised{2003}{1}{1}
\finalreceived{2003}{1}{1}

\begin{document}
\maketitle
\begin{summary}
  Although the notion of qualified names is popular in module systems, it
  causes severe complications. In this paper, we propose an alternative
to qualified names.
The key idea is to import the declarations in other modules
to the current module before they are used. In this way, all the
declarations can be accessed locally. However, this approach is not efficient in memory usage.
Our contribution is the {\it module weakening} scheme which allows us to import the minimal parts.
As an example of this approach,
we propose a module system for functional languages.
\end{summary}

\newcommand{\muprolog}{{Rec$^{\add,\adc}$}}

\renewcommand{\intp}{eval} 

\section{Introduction}\label{intro}

Modularity is the key technique for dealing with large programs.
Most modern languages (including object-oriented ones)
employ {\it qualified names} of the form $m.f$ to access
a method $f$ in a module $m$.
Although the notion of qualified names is easy to
implement, it leads to unnecessarily long names and it runs counter to the core of knowledgebase 
representation, i.e., the conciseness.
Thisl problem should be eliminated to
preserve clean and concise codes.
In order to do so,  we consider here an alternative
to qualified names.

The key idea is to import the declarations in $m$
to the current module before they are used. In this way, all the
declarations can be accessed locally.

To be specific, we propose to add the following:

\begin{itemize}
  

\item  import declarations  (MI declarations) of the form

 \[ /m \]
\noindent   where $/m$ is a module name.
   
\item  interactive/querying  declarations (MQ expressions) of the form (inspired by
   the  work  in \cite{Japfi})
\[(f(t_1,\ldots,t_n)=V)^{/m} \]
\noindent 
  where  $f(t_1,\ldots,t_n)= V$ is a query to a module $m$ and $V$ is a free variable not appearing in $x_1,\ldots,x_n$. 
\end{itemize}
  
\noindent 
The former one has the following  semantics: the 
declarations in $/m$ are  to be added to the
current program. 
This expression thus supports the idea of importing
{\it all}  the declarations of a module.

The latter one has the following  semantics:  first
evaluate $f(t_1,\ldots,t_n)$ w.r.t. $/m$ and set $V$ to the
resulting value $w$. This expression thus supports the idea of importing {\it some} (
{\it logical consequence} of)  declarations of a module. This is called {\it module weakening/module customization}.
Note that the notion of MQ declarations is a novel feature which is not
present in
traditional  languages. For example, $(fib(3)=V)^{mf}$ is a querying
declaration where the value of $v$ is not known. Its value is later
determined by evaluating $fib(3)$
w.r.t. the ``fibonacci'' module
$mf$.

Although our approach can be applied to other programming paradigms,
 this paper focus on  functional languages.
That is, we extend a   functional language with MI/MQ declarations. 

\section{A Case Study: Functional Languages}\label{func}

The theory of recursive functions, which we call $REC$,
provides a basis for functional programming.
It includes operations of composition, recursion, $etc$. Although $REC$ is quite expressive,
it does not contain local module mechanisms.

To fix this problem,  a modern way to add local declarations is to introduce
 declarations-implication expressions (DI expressions) of the form:
 
 \[ D \mli E \] 
\noindent where $D$ is a set of function declarations.
 This expression is adapted from the work in \cite{Mil89jlp}.

The above  has the following intended semantics: the MI/MQ
declarations in  $D$ are first  processed and then are  to be added to the
current program  in the course of evaluating $E$. 

This paper proposes an extension of the core functional languages with
DI/MI/MQ  expressions.

\section{Examples}

We assume a fibonacci  module $mf$ and a prime module $mp$ which
respectively contains the definitions of $fib(n)$ and $prime(n)$.

\begin{exmple}
  \% fibonacci module. \\
  \% $fib(n)$ returns $n$th Fibonacci number. \\
/mf = \% fibonacci module\\
$fib(1) = 1$.     \\
$fib(2) = 1$.     \\
$fib(n+2) = fib(n)+fib(n+1)$ \\
 \end{exmple}
\begin{exmple}
  \% module prime \\
\% prime(n) returns true if $n$ is prime \\
/mp =  \% prime module\\
$prime(n) = prime\_aux(n,n-1)$.     \\
$prime\_aux(X) = \ldots$.    
\end{exmple}

An  illustration of DI expressions
is provided by the following definition of the
function $primefib(n)$ which returns true if $n$th Fibonacci number is
prime:

\begin{exmple}
/mw =   \\
$      primefib(X)$ ${\rm =}$ \> \hspace{5em}   \\
  \> $/mf \mli /mp \mli prime(fib(n))$ 
\end{exmple} 
\noindent
The body of the definition above contains  DI expressions.
As an  example, evaluating $primefib(3)$ would result in
adding all the declarations in $mf,mp$, and then evaluating
$prime(fib(3))$. The machine returns true, as 2 is prime.

MQ declarations  are intended to add only some (logical consequences)
of a module to the current module.
  An
illustration of this  aspect is shown here.
\begin{exmple}
$      primefib(n)$ ${\rm =}$ \> \hspace{5em} $(fib(n)=v)^{/mf}$  \\
  \> $ \mli (prime(v)=w)^{/mp} \mli prime(fib(n))$ 
\end{exmple}
\noindent The body of the definition above contains  MQ declarations.
As a particular example, evaluating $primefib(3)$ would result in
adding only the two declarations $fib(3)=2$ and
$prime(2)= true$ to the program, and then evaluating  $prime(fib(3))$.
That is, $v$ is set to 2, $w$ is set to $true$.

\section{The Language}\label{s:logic}

The language is a version of the core functional languages 
 with DI/MI/MQ expressions. 
It is described
by $E$- and $D$-rules given by the abstract syntax as follows:
\begin{exmple}
  \>$E ::=$ \>  $c \sep x \sep  h(E,\ldots,E) \sep     D \mli E \sep T$ \\
 \>$D ::=$ \>  $ /m \sep f(t_1,\ldots,t_n) = E   \sep (f(t_1,\ldots,t_n) = v)^{/m}
 \sep$\\
 \>\> $D \land D   $\\
\end{exmple}
\noindent
In the abstract syntax, $E$ and $D$ denote the expressions and the definitions, respectively.
In the rules above, $c$ is a constant, $x,v$ are variables, $t$ is a term which is either a variable or a constant, and
$m$ is a module name. A set of function definitions $D$
is called a program in this language.

\newcommand{\bc}{bc}

We will present the semantics of this language in the
style of \cite{KK07}. It consists of two steps.
The first step is to preprocess and instantiate
all the querying declarations in $D$
by invoking queries to their corresponding module.
The second step is described as a set of rules in Definition 1.
The evaluation strategy assumed by these rules is an eager evaluation. 
 Note that execution  alternates between 
two phases: the evaluation phase defined by \textit{eval}
and the backchaining phase by \textit{bc}. 

In  the evaluation phase, denoted by $\intp(D,E,K)$, the machine tries to evaluate an expression $E$ from the program $D$, a set of definitions, to get a value $K$.  Note that these rules written in logic-programming style, \ie, $\intp(D,E,K)$ is true if the evaluation result of $E$ in $D$ is $K$. 
For instance, if $E$ is a function call $h$, the machine first evaluates all of its arguments and then looks for a definition of $h$ in the program in the backchaining mode.


The rules (1) -- (4) describe the backchaining mode, denoted by $bc(D_1,D,h,K)$.
In the backchaining mode, the machine tries 
to evaluate a function call $h$
by using the function definition in the program $D_1$.

\begin{defn}\label{def:semantics}
Let $E$ be an expression  and let $D$ be a program.
Then the notion of   evaluating $\lb D,E\rb$ to a value $K$ --- $\intp(D,E,K)$ --- 
 is defined as follows:

\begin{numberedlist}


\item    $\bc(h(c_1,\ldots,c_n) = E, D, h(c_1,\ldots,c_n), K)$ \\ if 
  $\intp(D, E, K)$. \% switch to evaluation mode.


\item    $\bc(D_1\land D_2,D,h(c_1,\ldots,c_n),K)$  \\
 if   $\bc(D_1,D,h(c_1,\ldots,c_n),K)$. \% look for $h$ in $D_1$

\item    $\bc(D_1\land D_2,D,h(c_1,\ldots,c_n),K)$  \\
  if   $\bc(D_2,D,h(c_1,\ldots,c_n),K)$. \% look for $h$ in $D_2$


\item    $\bc(h(x_1,\ldots,x_n) = E, D, h(c_1,\ldots,c_n),K)$ \\
if   $\bc(h(c_1/x_1,\ldots,c_n/x_n) = E', D, h(c_1,\ldots,c_n),K)$ where
    $E' =   [c_1/x_1,\ldots,c_n/x_n]E$. \% argument passing to
 $h$ and $E$.

\item    $\intp(D,\top,\top)$. \% 
 $\top$ is always a success.

\item   $\intp(D, c, c)$.  \%   success if $c$ is a constant.

\item    $\intp(D,h(c_1,\ldots,c_n),K)$ \\ if   $\bc(D,D,h(c_1,\ldots,c_n),K )$. \% 
 switch to backchaining by making a copy of $D$ for a function call.

\item    $\intp(D,h(E_1,\ldots,E_n),K)$ \\ if $\intp(D,E_i,c_i)$ and $\intp(D,h(c_1,\ldots,c_n),K)$.
 \%  evaluate the arguments first.


\item $\intp(D,D_1\mli E,K)$ \\ if  
$\intp(D\mlc D_1,E,K)$
\%  DI expressions. 

\item $\intp(D,D_1\mlc/m\mlc D_2\mli E,K)$ \\ if  
  $\intp(D, D_1\mlc D_3\mlc D_2 \mli E,K)$, provided that $D_3$ is the
  declarations contained in module $m$.
\%  MI expressions. 

  \item $\intp(D, D_1\mlc (f(t_1,\ldots,t_n)=v)^m \mlc D_2\ \mli\ E,K)$ \\ if  
  $\intp(D, D_1\mlc(f(t_1,\ldots,t_n)= w)\mlc D_2\mli E[w/v],K)$, provided that $w$ is the
    value obtained by evaluating $f(t_1,\ldots,t_n)$ w.r.t. the module $m$.
 \%  MQ expressions.

\end{numberedlist}
\end{defn}

\noindent  Note that, in rule (11), a module that is queried could itself query other modules.
For simplicity, other popular constructs such as \textit{if-then-else} and pattern matching are not shown above.
 If $\intp(D,E,K)$ has no derivation, it returns a failure.

\section{Conclusion}\label{sec:conc}

In this paper, we proposed an extension to functional languages with  
DI/MI/MQ  expressions. 
These expressions are 
particularly useful for avoiding qualified names in functional languages.

The MQ expressions  can be implemented by treating them like exceptions.
That is, when an MQ expression is encountered,
suspend the current execution, switch to another execution (
That is, evaluating a query with respect to another module) and then resume the
suspended execution.

Our ultimate interest is to design a module system for
Computability Logic \cite{Jap0}--\cite{JapCL12}.



\bibliographystyle{ieicetr}


\end{document}